\documentclass[runningheads]{llncs}
\usepackage{graphicx}
\usepackage[export]{adjustbox}
\begin{document}
\title{Fodor and Pylyshyn's Critique of Connectionism and the Brain as Basis of the Mind}
\titlerunning{Brain as Basis of Mind}
%
\author{Christoph von der Malsburg\inst{1}}
%
%
\institute{Frankfurt Institute for Advanced Studies, Frankfurt am Main, Germany \and
Institute for Neuroinformatics, ETH Zurich, Switzerland \and 
ZHAW Centre for Artificial Intelligence, Winterthur, Switzerland}

\maketitle              
\begin{abstract}
To this day there is no satisfactory answer to the question
  how mental patterns correspond to physical states of our brain.
For more than six decades, progress has been held up by the logjam
  between two traditions, one inspired by neuroscience, the other by digital computing.
This logjam is well illuminated by Fodor and Pylyshyn's article of 1988,
  which is mainly devoted to a critique of what they call Connectionism,
  but also lays bare weaknesses of the Classical approach which they defend.
As recent machine learning breakthroughs may be expected to through new light on the issue,
  it seems time to arrive at a synthesis of the connectionist neural approach
  and the classical stance based on symbol processing.
I will present and discuss an attempt at such synthesis
  in the form of structured, self-organized neural networks.

\end{abstract}
\section{Introduction: The Question}

Let me adopt, without any discussion, the attitude of monism:
  the conviction that mental phenomena are physical,
  that they are to be identified with complex patterns and sequences of patterns
  constituted by the building elements that we find in our nervous system.
What these physical patterns are and how they relate to mental patterns
  is a question of deepest importance.
Recent achievements of machine learning and artificial intelligence (AI) \cite{attention2017,gpt4}
  can be seen as realization of mind-like processes on an electronic basis.
These systems still have deficits in comparison to the human mind
  ---they lack own drives and the ability to learn directly from the environment---,
  but what they do achieve may be seen as essential breakthrough
  on the way to understand and replicate mechanism of the mind,
  as proof of principle: that mind is possible on a physical basis.
  
Those systems should, therefore, be a convenient opportunity to arrive at an answer
  to the question just raised, how mental patterns might be realized as physical patterns.
They offer something extremely valuable:
  a clear separation of form from content.
The form is expressed in terms of surprisingly concise programs or
  verbal and mathematical descriptions
  that specify a data framework, a basic mechanism of operation
  and a process of learning (back-propagation of error) from training material,
  while the content is treated as data and is exclusively provided by that training material.
  
The answer to our question accordingly should be contained in
  the concise descriptions of what I just called data framework and basic mechanism of operation.
Those descriptions speak of vectors and matrices (arrays of floating point numbers)
  and their algebraic concatenations in arrangements of layers and projections between these.
A decisive breakthrough in comparison to multi-layer perceptrons \cite{AlexNetNIPS2012},
  which still had severe functional limitations,
  was achieved by what is called attention
  and the transition to a generative mode \cite{attention2017}.
This is a mechanism that takes in arrays of input elements
  (language tokens or image patches expressed as `embedding vectors'),
  dynamically identifies from among them
  sets of elements that bear regular relations to each other
  and combines part of their content into new representations
  that are then handed on to the next layer.
This data framework is initiated with random numbers
  and in a process of extensive training it gradually absorbs
  structure from masses of human-generated material.

One now could entertain the hope that by scrutinizing the data frameworks and procedures
  that are the basis for those successful AI systems
  one could recognize the characteristic traits that
  after centuries of philosophy and research
  we came to associate with the working of the mind.
Not an easy task, even if that hope was justified,
  as all those systems have been developed in the spirit of engineering
  with very little regard to science \cite{BitterLesson}.
The task is made especially difficult due to a very fundamental deviation
  of machine learning from the mechanisms by which humans and animals acquire structure and knowledge.
Whereas those come equipped at birth with complete though simple functionality
  and pick up structure and knowledge by direct interaction with the environment,
  machine learning has to rely on masses of human-structured data and passively extract
  regularities from them by statistical means.
  
One consequently should perhaps see in the fundamental algorithms of machine learning
  more a function of copying devices and in the search for physically represented mental patterns
  one should rather scrutinize the structures that emerge in those systems by learning.
Unfortunately, when inspecting the vast matrices of trained systems they
  cannot be distinguished from random noise \cite{wolfram2023chatgpt},
  and attempts of identifying structural regularities, e.g., 
  \cite{chefer2021generic,aflalo2022vlinterpret,hao2021selfattention}
  bring up very little useful insight for our purpose.
Thus we are, at least for the time being, in the ironic situation
  that we have a mind in a bottle but cannot make sense of it.
Let us therefore look what insights we can gain from the more scientifically minded
  traditional attempts at AI.

\section{Two Camps: Symbols vs. Neurons}

It may sharpen the question we are posing here by going back to the old controversy
  between two approaches towards AI and the attempt to understand the mind on a physical basis.
Like in a seewsaw, two communities of researchers took turns in dominating the discussion,
  repeatedly sending the other camp into a `Neural Winter' or an `AI Winter'.
  
One camp took off with a workshop in 1956 at Dartmouth College \cite{moor2006dartmouth}.
That camp secured for itself the name Artificial Intelligence
  (now sometimes referred to as classical or good old AI, GOFAI),
  so let us refer to this camp as ``Classical AI". 
The movement was energized by the early availability of computers
  as basis for the explicit modeling of cognitive processes
  by handling parsimonious symbols according to fixed rules.

The other approach, inspired by the nervous system, 
  restricts itself to considering models in the form of neural networks
  and is dominated by two concepts.
One of them has been formulated by D.O. Hebb and F. Hayek \cite{hebb,hayek}
  who proposed (not without precedent) the associative idea
  ---neurons that fire repeatedly together strengthen their excitatory connections---,
  and who proposed sets (``assemblies", ``pools") of mutually supporting neurons as
  realization of composite mind entities\footnote{
  (assemblies and their associative formation was later variously formalized as ``associative memory"
  \cite{steinbuch1961lernmatrix,willshaw1969non,Palm1980,Hopfield2554,cohen1983absolute}.}.
The other concept was proposed by Frank Rosenblatt in the form of the multi-layer perceptron \cite{rosenblatt},
  which is based on feed-forward hierarchies of feature detectors
  and is trainable by back-propagation of error \cite{rosenblatt}, p. 272
  (whichlatter received a clear mathematical formulation only later, e.g. \cite{PDP-backprop}).

Each of the two approaches has much to say in its own favor and against the other.
The two approaches and their strengths and weaknesses are best described relative
  to the four architectural questions that must be answered to specify the working of mind and brain:
\begin{itemize} 
\item Q1 What is the physical nature of the mental patterns
  that are active moment-to-moment in the system?

\item Q2 What kind of variables describe the permanent structure of the system?

\item Q3 What is the mechanism of creating and transforming mental patterns?

\item Q4 How are the structural variables modified during learning and self-organization?
\end{itemize}
The first architectural question is the focus of this communication,
  but it can hardly be understood without considering the other three.
The two camps  give different answers to these questions, see the next section,
and up until now the situation must be seen as a stand-off with no accepted synthesis in sight.
In view of recent AI breakthroughs it may be time to pick up the discussion
  and attempt this necessary synthesis.

\section{Fodor and Pylyshyn's Critical Analysis}

The controversy between the classical AI and the neural approach
  was most pointedly formulated in a paper \cite{FodorPylyshyn} 
  published in 1988 by Jerry Alan Fodor (1935- 2017) and Zenon Walter Pylyshyn (1937–2022),
  two American resp. Canadian philosophers and cognitive scientists.
This paper (referred to here as F\&P) may be a good starting point to bring the two opposing perspectives together.
F\&P appeared at the height of the neural camp's euphoria of the eighties,
  took the position of what they call `Classical theory' and set it against `Connectionist theory'.
They see both as cognitive architectures that postulate the existence
  of representational mental states (as answer to our architectural question Q1).
Couched in terms of answers to those architectural questions,
  F\&P's characterization of the two approaches are:
  
\vspace{3mm}
\noindent The Classical approach according to F\&P:
\begin{itemize}
\item  Q1: Mental states are strings of symbols.
  These have the form of `atoms' (structure-less tokens) or `molecules': expressions composed of atoms.
\item Q2: The structure of the system is a set of rules, encapsulated in a computer program.
\item Q3: Mental states (strings of symbols) are generated and processed by the execution of rules
  as implemented in a Turing machine.
  The rules are sensitive to the compositional structure of expressions.
  Compositional structure (abstracting from the identity of constituent atoms)
    induces `symmetries', relations between different expressions that share compositional structure.
  This is the source of `systematicity' and `productivity',
    the ability of the mind to produce and deal with new expressions on the basis
    of common compositional structure.
\item Q4: Not a single word on the question of how rules or programs arise.
\end{itemize}
The Connectionist approach according to F\&P:
\begin{itemize}
\item Q1: Mental states are sets of active neurons.
  These may carry labels (indicators of type),
  but these labels exist only in the mind of the spectator,
  may play a role during the wiring-up of the system (Q4), but
  have no influence on processing (Q3).
\item Q2: The structure of the system is constituted by the existence and strenghts of excitatory  connections
  (presumably also inhibitory connections, although they are not mentioned).
  Connections are restricted to a causal role ---switching neurons on or off---
  and cannot be distinguished in terms of any kind of type.
\item Q3: Mental states ---the sets of active neurons---
  are generated and processed by the exchange of excitation (and inhibition).
\item Q4: Neural connections ---the parameters constituting the structure of the system---
  are generated and modified by an associative mechanism
  that is sensitive to the statistics of neural firing.
\end{itemize}

\subsection{F\&P's criticism of the Connectionist approach}

Their criticism is directed at one single point: the insensitity of neural systems to
  compositional structure and the ensuing absence of systematicity and productivity.
They most concretely illustrate their point by the discussion of an example,
  the sentence `John loves Mary', going through variations of how this sentence
  might be represented as a set of active neurons.
A set \{John; loves; Mary\} (where each element stands for a labelled neuron)
  doesn't do, as it could be confuced with the sentence `Mary loves John'.
Also sets of the form \{John-actor; loves; Mary-patient\} are to be rejected because
  specialized nodes such as `Mary-patient' don't generalize well,
  as a state \{Mary-actor; loves; John-patient\} would be seen as
  totally unrelated to \{John-actor; loves; Mary-patient\} because `Mary-patient' and `Mary-actor'
  would be separate neurons sharing only part of their label which,
  however, is not represented in the system but only in the spectator's head.
F\&P argue further against that same `solution': {\it ``Consider a system in
which the mental representation that is entertained when one believes
that John loves Mary is the feature set \{John-actor; loves;
Mary-patient\}. What representation corresponds to the belief that
John loves Mary and Bill hates Sally? Suppose, pursuant to the present
proposal, that it’s the set \{John-agent; loves; Mary-patient; Bill
agent; hates; Sally-patient\}. We now have the problem of
distinguishing that belief from the belief that `John loves Sally and
Bill hates Mary'; and from the belief that `John hates Mary and Bill
loves Sally'; and from the belief `John hates Mary and Sally and Bill
loves Mary' etc. since these other beliefs will all correspond to
precisely the same set of features. The problem is, of course, that
nothing in the representation of Mary as Mary-patient specifies
whether it’s the loving or the hating that she is the patient of;
similarly, mutatis mutandis, with the representation of John as
`John-actor'."}  This ambiguity could be avoided, F\&P point out, with
neurons standing for whole sentences, like
'John-agent-hates-Mary-object', but rightly reject this as utter
absurdity.

With these examples F\&P illustrate the impossibility to let
unstructured sets of neurons express compositionality without becoming ambiguous.
(This is what has become to be called the binding problem \cite{Report,AmIThinking,roskies1999binding,burwick2014binding}.

F\&P add a brief discussion of graphs composed of neurons and links
  as mental representation but reject it,
  as the connections between neurons serve the sole function of
  transmitting excitation and cannot be interpreted as typed arrows.

\subsection{Comments to F\&P's perspective}

I repeat that in my view F\&P offers a fair and insightful characterization
  of the positions of the Classical and Connectionist camps.
This is particularly valuable as they are {\it volens nolens} laying bare
  grave conceptual deficiencies on both sides.
At the time there were no full-scale artificial systems that could seriously be recognized as models of the mind 
  and the discussion was all based on extrapolation from very simple examples.
Both sides were anthropomorphizing these examples,
  labeling the symbols or neurons so that they could take on meaning in the observing human's mind,
  whereas in reality they would have to acquire their meaning
  by way of their integration into a full functional system.

F\&P's pointing out of the importance of compositional structure and
  sensitivity of processes to this structure, the central point of the paper,
  is of course of great importance for understanding the mind
  and a valid criticism of the Connectionist approach at the time.
They, and the Classical camp, can't, however, claim to have solved the problem.
Recognizing and exploiting the relation between the sentences
  `John loves Mary' and `Peter hates Carl'
  necessitates recognition that `John', `Mary' etc. are persons,
  that `loves' and that `hates' are interpersonal relations
  in order to make the sentences comparable.
This all is recognized and represented in the (anthropomorphizing) reader's mind,
  but for the imagined system to function it has to happen {\it within} it.

An obvious solution would be representation of all such attributes by what F\&P call
  micro-features attached to the surface symbols.
This, however, they explicitly reject, insisting for symbols to speak for themselves
  (from which one may conclude that they interpret `language of thought' literally,
  as linear sequence of atomic symbols).
Thus, the task of recognizing and representing the necessary attributes of symbols
  would fall, in the Classical view, to the rules that are to handle those symbols.
This would be acceptable if the Classical view offered mechanisms
  by which the rules could arise (our architectural question Q4),
  but on this question it has a blind spot 
  (as has formal mathematics).

F\&P are not explicit at all with regard to mechanisms that could exploit compositionality
  for the purpose of systematicity and productivity,
  neither on the Classical nor on the Connectionist side.
Presumably, the representation of `John loves Mary' would have to contain some
  substructure of the kind `sentient entity bears an emotional relationship to another sentient entity'.
This substructure could then serve as template to form other expressions of similar meaning.
On the Classical side, this template would have the form of a rule that has the
  ability to recognize or form sentences that conform to this pattern.
For that rule to recognize `John' as sentient entity or,
  on the productive side, to find other sentient entities,
  a micro-feature of that kind would have to be attached somehow to `John' and analog entities,
  and that requirement would have to apply, of course, both on the Classical
  symbol-processing and the Connectionist neural side.
(To afford productivity, the Classical side would have in addition to provide for a mechanism,
  once a sentence of the type `John loves Mary' has somehow been formed,
  to create the corresponding rule, as answer to Q4.)

We have to conclude that ``atomic" symbols, like `John' or `loves',
  must be equipped with a rich set of micro-features.
On the neural side, where these micro-features would be represented by neurons,
  this requires a solution to the binding problem,
  in this case a means to express the fact that all these features relate to each other
  and together form a composite symbol.

Within a composite expression like `John loves Mary',
  in which all ``atoms" come equipped with a rich collection of micro-features,
  it will then be possible to find a substructure that it shares with
  other composite expressions like `Bill hates Sally', expressions that are,
  in the sense of containing this identical substructure, analogous to each other.
Shared substructure is the true basis of the productivity and systematicity
  that are at the center of F\&P's argument.

At the time of F\&P the heated discussion between the two camps was based
  on mental extrapolation from pathetically small demonstration pieces,
  but 30 years after F\&P a wave of full-scale functional systems has sprung up,
  beginning perhaps with \cite{attention2017}.
The essence of these systems is, I believe, captured by the last paragraph.
They run on essentially conventional computers,
  structures are compositional and processes are sensitive to composition,
  systematicity is exploited to full extent for the benefit of productivity
  (as demonstrated specifically by translating text from one language into another
  on the basis of structural analogies).
Can one say, looking at the intervening period, that the battle 
  between the Classical and Connectionist camps has been won
  by one  of them?
The historical roots of those Transformer systems are
  the multi-layer perceptron \cite{rosenblatt,fukushima1980neocognitron}
  (which are part of the Connectionist approach)
  and the embedding vectors (sets of micro-features) go back to the field of
  Natural Language Processing \cite{harris1954distributional}.
  (which probably should be counted as part of the Classical approach).
So one may perhaps speak of a synthesis of ideas from both camps.

The true winner, for the moment, is Engineering and Industry and
  the academic world ---cognitive science, neuroscience and all---
   is temporarily side-lined.
Is that the end of the story?
There are some fundamental differences between the machine learning systems
  and natural brain and mind,
  among them the very different mode of learning,
  and there are fundamental questions
  concerning, for instance, the status of emergence. 
I would therefore like to complement my discussion of the physical nature of mental states
  by a summary of a neural cognitive architecture I have developed over the decades,
  as basis for discussing those and other open questions.

\section{Mind States as Represented by Structured Nets}

\subsection{Neural Nets}

Consider cortical areas and their interaction
  through dynamic fiber projections.
Cortical areas are two-dimensional sheets of neurons.
As result of early ontogenesis, neurons within areas have at their disposal 
  a large number of short-range excitatory connections
  each of which can choose, in the course of development,
  among a number of target neurons.
Let me first give answers to the two architectural questions,
  concerning the neural nature of mental patterns (Q1) and
  the mechanism of establishing them (Q3).

My central claim is that mental patterns are neurally represented by what I call neural nets or simply nets.
A net has the form of a specifically connected sparse set of currently active neurons
  (sparsity meaning that only a very small fraction of all neurons is active in a net, say, one in $10^4$ neurons).
The condition for a neuron to fire in a net is that it is supported by
  a minimum number of (say, a few dozen) active excitatory connections, that is,
  excitatory connections from other neurons that are active in the same net%
\footnote{According to \cite{Abeles} it takes 10 simultaneous EPSPs for a neuron to reach threshold.
  As precise simultaneity may be unrealistic, a few dozen may be needed.}.
  
Meeting this condition of a minimum number of active connections
  for all neurons in a net is not trivial
  and requires very specific connectivity patterns.
Assuming the connectivity were random,
  a fraction of $10^{-4}$ neurons were active, and a neuron had $10^4$ excitatory connections, 
  the expected number of active connections would be 1.
Accordingly it needs a specific developmental process to form nets,
  i.e., to form sets of neurons each of which has the required minimum
  number of connections from other members of the same set.

How can nets that have been formed be activated in a cortical area?
This needs a two-stage process.
In stage one, afferents to the cortical area activate a large number of neurons
  (a fraction $10^{-3}$ or even $10^{-2}$ of all neurons) transiently%
\footnote{Afferents have privileged access to neurons, connecting with their synapses near to the
  spike-triggering axon hillock, and are not subject to the above minimum number condition.}.
Stage one is ended by a wave of inhibition (perhaps the balancing inhibition described in the literature \cite{Vogels}),
  which silences all neurons except those that are supported by the required minimum number of excitatory  signals
  from other neurons that had also been activated within the stage-one transient.
Stage two amounts to the projection of a net out of the cloud of afferently activated neurons.

A neuron together with the set of neurons that support it in a net
  is called a {\it minimal fragment}.
A minimal fragment is composed of a central neuron together with its supporting neurons.
A net thus is composed of as many minimal fragments as there are neurons in it
  and conforms to the stringent condition that all these fragments fit together,
  that is, all supporting neurons are themselves also central neurons of fragments.

Besides minimal fragments one may speak of a hierarchy of larger fragments,
  sets of neurons with mutual support which have a high likelihood of being activated together.
Smaller fragments are nested in larger ones, thus forming a hierarchy.
This hierarchy replaces, to a large extent, the hierarchy of feature types 
  in multi-layer perceptrons such as \cite{AlexNetNIPS2012}.
Beyond a certain size, nets tend to be unique, being activated only once in a lifetime,
  but are throughout materialized and covered by net fragments.
Net fragments cannot subsist on their own and can only be activated as part of larger nets.

As a cortical neuron has $10^4$ incoming connections \cite{Gulyas2001}
  and the minimum number of necessary active incoming connections may be less than 100,
  a neuron can be part of quite a number of different fragments,
  some of which may overlap in more than the central neuron
  while some may be disjoint and overlap just in this (central) neuron.
This participation of a neuron in multiple fragments has important
  consequences for the assignment of meaning to neurons, see below.

\subsection{Communication between Nets in Different Cortical Areas}

Cortical areas are interconnected in the form of a hierarchy \cite{Felleman91}.
They are coupled through axonal fiber connections up and down the hierarchy, 
  see for instance \cite{Hausser23}.
On the way up one speaks of afferents, on the way down of top-down connections.
The connections between cortical areas are very sparse in comparison to the within-area connections
  and cannot transfer much content from one area to another,
  only exert enough influence to help select between nets
  that are constituted by intra-area connections.
  
Let me discuss the step from a lower, more peripheral, area
  to a higher, more central one, as a step of abstraction.
Lower levels typically are subject to variation within sensory or motor spaces that are affected by
  the position of the body relative to the external world,
  whereas higher levels tend to relate to intrinsic structure of the environment.
Thus, there are multiple nets on lower levels that should all relate to and select
  a single net at the next-higher area.
The relationship between corresponding nets (i.e., nets that are relating to the same external structure)
  on different levels of the hierarchy is one of homeomorphy%
\footnote{the term homeomorphy is borrowed from the mathematical field of topology.}:
  two nets are related by homeomorphy if there is a one-to-one mapping between them
  that maps neurons connected in one net onto neurons that are connected in the other net.

Maps between homeomorphic nets in different cortical areas
  are to be realized by axonal projection fibers in both directions.
In order to support rapid processing and accommodate many different mappings,
  all necessary projection fibers must be physically present%
\footnote{Shifter circuits \cite{Anderson1987} keep the number of required fibers low 
\cite{WolfrumMalsburg2007}.}.
Of these, a small subset is to be selected to form a particular mapping.

Let me discuss the case of vision, for which this selection of projection fibers
  is the most demanding.
A visual input to primary visual cortex
  excites to two-dimensional distribution neurons of different feature types.
(Feature type refers to the shape of the receptive field of a neuron
  that lets it respond to a particular local visual pattern.)
Different nets that correspond to the same visual scene or object
  seen under different eye position, that is, in different retinal coordinates,
  can be mapped to each other under preservation of neighborhood relations and feature type.
All these lower-level nets are to be represented in a higher cortical area
  (presumably in infero-temporal cortex \cite{tanaka1996inferotemporal}) by the same net.
The mappings from the different nets corresponding to the same visual pattern
  thus share two properties, preservation of topology and of feature type.

Consider any pair of local patches,
  one within primary visual cortex and one within a higher-level area.
Any such pair is connected by a mapping 
  (a `maplet' \cite{zhu2004maplets}) 
  that connects neurons one-to-one in a neighborhood- and neural type-preserving fashion.
Different pairs of nets that happen to fall with corresponding regions onto such pair
  make active use of different subsets of the  fibers that make up the maplet
  connecting the two patches. 
It would be desirable to let these fibers collaborate and switch each other on and off
  (as do neurons in a net fragment).
Such collaboration between fibers of the same maplet has indeed been proposed
  as mediated by {\it control units} \cite{Anderson1987}.
Control units gate open a set of fibers if the pattern of activity
  carried by the fibers is strongly correlated with the pattern of activity
  in the fibers' target neurons \cite{WoWoLuMa}.
The different control units whose maplets together form a
  coherent topological mapping can excite each other,
  so that a whole mapping can be switched on efficiently in a collective action,
  while control units that are inconsistent with each other
  (like mapping different patches of the lower level to the same patch of the higher) inhibit each other.
  
\subsection{Development: Learning, Self-Organization of Nets}

How does the connectivity that supports net structure develop through learning and self-organization?
A requirement is that the sought-for final connectivity must be accessible 
  within the range of potential modifications to the initial connectivity state.
Ontogenesis under genetic control generates the initial connectivity of the brain,
  which thus defines the hull within which connectivity develops for the rest of life.
Specifically, the visual modality is dominated by the topology of two-dimensional image domains ---
  neighborhood in retinal, in object-centered or in scene-centered coordinates.
Correspondingly, the initially stochastic connectivity must support
  interactions over the lateral range of pattern correlations encountered in visual signals,
  and plasticity mechanisms must provide for enough flexibility
  so that the final connectivity can be reached.

The task of the plasticity mechanisms of the brain (architectural question Q4)
  is to generate the connectivity (answer to Q2)
  that supports the mechanism (Q3) that generates activity states (Q1).
In these states, the active neurons are stabilized by afferent and top-down signals and
  by mutual excitatory support within cortical areas.
The mechanism (Q4) to generate the structural basis for such states
  relies on Hebbian synaptic plasticity \cite{hebb}.
According to this mechanism,
  excitatory synaptic connections are formed, strengthened and stabilized
  by the success of the signals they carry in predicting activity in their target neuron.
The mechanism of intra-areal network self-organization is thus a clear-cut example of emergence:
  there are certain connectivity patterns that have the ability of self-stabilization
  (all participating neurons are stabilized in their activity by lateral support
  and individual connections are successful in predicting the activity of their target neuron)
  and these networks thereby have the ability to grow out of the dynamic turmoil of the system.

\subsubsection*{Development of Intra-Areal Nets.}

Initially, when the lateral connections within a cortical area are still random,
  Hebbian plasticity is driven exclusively by the signal statistics of afferent fibers,
  and pairs of neurons that are frequently co-activated by the afferent input
  may develop a connection.
Once a substantial number of such lateral connections have formed,
  the level of activation of a cortical neuron in response to afferent input
  can be graded according to the level of lateral support it gets
  from other neurons activated by the same input.
A coherent net can thus grow by sequestering neurons and
  establishing connections between them.
Each larger afferent pattern to an area is unique and never repeats.
But there may be relatively small subsets of fibers,
  activity on which is dominated by repeating patterns.
(This is, for instance, the case in primary visual cortex
  for sets of sensory fibers that come from local patches in the retinae.)
These patterns can then be picked up by synaptic plasticity,
  so that different patterns get represented by different net fragments.

To keep the connectivity in nets sparse,
  neurons accept only a limited number of active connections
  (connections from other active neurons within the same state),
  so that the required minimal number is not exceeded sustantially
  and synapses have to compete within a net.
(Connections to a neuron that are active in different net fragments the neuron belongs to
  don't compete with each other.)

\subsubsection*{Development of Inter-Areal Connections.}

The connections between two cortical areas $A$ and $B$ serve to
  implement homeomorphic mappings:
  connectivity-preserving one-to-one connections
  between active neurons in $A$ and active neurons in $B$.
Let's assume that in both areas all nets are constituted by short lateral connections.
Let there be two active patterns, one in $A$, one in $B$.
Then according to network self-organization \cite{so2,zornetzer-two}
  plasticity will select inter-areal projection fibers so
  that neighboring neurons in one area project one-to-one to neighboring neurons in the other.
This mechanism has been intensively studied in the context of the ontogenesis
  of fiber projections from the retina to the optic tectum \cite{rettec,rettecReview}
  and has the particular property of `systems matching' \cite{gaze1972visual},
  in our case meaning that the whole net in $A$ projects to the whole net in $B$.

As remarked above, an important application is that different nets in area $A$
  are homeomorphic to the same net in area $B$.
This applies, for instance, when different images, produced under eye motion
  during inspection of an object and represented by different nets
  in primary visual cortex (taking the role of our area $A$)
  are to be represented invariantly in another cortical area
  (in infero-temporal cortex, our area $B$ for the present purposes).
Inter-areal connections between $A$ and $B$ will then have to develop
  to realize homeomorphic patterns between those different nets in $A$
  and the one net in $B$.
Control units \cite{Anderson1987,WoWoLuMa} that initially are uncommitted to what fibers they control
  can then home in in a kind of EM process \cite{Malsburg,EM2008} to connect to specific sets of fibers
  that are again and again active together.
This process has been demonstrated in \cite{Fernandes}%
\footnote{and can be extended to multi-layer shifter circuits \cite{WolfrumMalsburg2008}.}.

\subsection{Interpretation of Nets as Mental Patterns}

How can nets be interpreted as mental patterns and as the language of thought,
  how can they bridge the conceptual chasm between the camps of classical AI and neural theory?
In the eyes of F\&P and classical AI,
  handling of cognitive entities as a whole
  according to rules and with sensitivity to compositional structure
  is crucial for the realization of intelligence
  while the neural camp had yet to come up with mechanisms
  that could show these capabilities.
The neural camp countered with the argument that
  the rules for handling symbols had to fall from heaven
  instead of being learned from example or derived from first principles,
  and that symbols were mere structure-less tokens that
  failed to render the fine-grained content of entities
  (beyond the level of detail implied by compositionality).

How do nets and net fragments relate to the two camps
  and how do they solve those problems?
They are composed of neurons and can represent whatever they stand for in detail,
  rendering, for instance, the image of an object in primary visual cortex
  with high resolution in terms of various qualities such as
  texture, color, surface shape or motion.
On the other hand, nets are integral entities that are held together by
  overlap and mutual support of fragments.
Nets solve the binding problem that was implicitly posed by F\&P and in \cite{Report},
  by representing composite entities by structured nets
  that can interact with other nets only under structural constraints.
Decisions are met by whole nets, not by individual neurons,
  as neurons can be stably active only under support from a net.

\begin{figure}[h]
\includegraphics[width=0.5\textwidth, center]{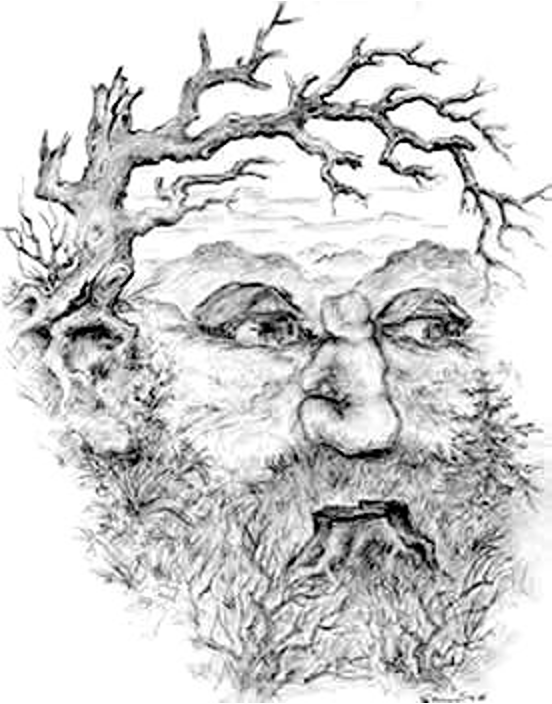}
\caption{Recognition of the face as global Gestalt leads to interpretation 
  of local detail in the light of the whole
  (e.g., the sitting person as a nose).
  A net that responds to the face as a whole
  supports only those local features and fragments that fit in with the whole,
  while feed-forward feature detectors would be subject to the fallacy of early commitment
  \cite{Marr} and decide on merely local information.}
  \label{fig1}
\end{figure}

This aspect is crucial and well illustrated in the Gestalt phenomenon \cite{wertheimer-gestalt}:
  the ability of our visual system to extract and recognize
  a coherent form even if its elements are immersed in
  locally misleading detail (see the figure).
The Gestalt rules that are invoked by that movement \cite{wertheimer-gestalt}
  to explain the phenomenon
  are naturally implemented by net fragments,
  whose activation depends not only on the local presence of afferent features
  but also on overlap with neighboring fragments and global coherence.

Neural assemblies \cite{hebb,Hopfield2554} can represent detail and can
  be formed by synaptic plasticity \cite{Hopfield2554,cohen1983absolute}, but
  different assemblies cannot coexist in a cortical area without 
  merging into one (the superposition catastrophe \cite{AmIThinking}). 
Avoiding this by allowing only one assembly per cortical area \cite{Papadimitriou20}
  is too narrow a constraint for a cognitive architecture.

Mental representation of F\&P's central example, the sentence `John loves Mary',
  has to contain, first, quite a range of elements giving detail to the
  sentence's constituents, `John', `loves', `Mary', second,
  it has to contain composite abstract entities of the kind `subject-verb-object'
  and, third, has to specify the relations John-subject, loves-verb and Mary-object.
The claim here is that this composite structure and its functional integration
  with all those ideas and actions
  that are evoked by the sentence in a listener's mind
  are naturally represented by one coherent net
  that is coherently composed of fragments that have been formed and learned
  in previous contexts.

An important theme in philosophy and cognition is schema theory 
\cite{Kant,Piaget,Bartlett,rumelhart2017schemata}.
According to schema theory general concepts are represented by abstract schemata
  which can be recognized in concrete instances and which can help
  to construct new instances.
In the context of F\&P, representation and handling of a composite entity
  like a sentence requires reference to a schema
  that represents its abstract structure and thus can relate the sentence to others
  that conform to the same schema.
In terms of nets and their interactions,
  concrete instances are represented in a lower cortical area
  and schemata are represented in a higher area.
The schema-instance relationship is established by a homeomorphic fiber projection between the areas,  
  in which a subset of the neurons in the instance's representation in the lower area
  link to neurons in the schema representation on the higher area
  under preservation of type and connectivity.
Net theory is, for the first time, offering a mechanism that promises
  to be able to activate a schema in response to an instance
  (and presumably to create instances in response to the activation of several schemata
  that complement each other in terms of mutually abstract patterns).
This has been demonstrated on the particular problem of relating
  facial images to a face schema \cite{WoWoLuMa}.

If the physical implementation of mental patterns has the form of nets,
  how are nets to be interpreted in terms of meaning?
When looking at a TV screen through a hole in a sheet of paper so that only a few pixels are visible,
  all there is by way of meaning is a description of the pixels as bright or dark or colored.
When taking away the sheet of paper, those pixels take on very specific meaning
  as part of objects or surfaces.
Analogously, neurons in the brain acquire meaning by their integration into the
  larger context provided by a net.
Some neurons that are, in some instant, activated in primary visual cortex by signals
  coming from the tip of the nose of a person's face acquire meaning from
  becoming connected by projection fibers to a general face schema higher up
  and to a representation of the particular person even higher up,
  projections that are only possible due to the integration of the
  tip-of-the-nose neurons into their lateral net context
  that projects homeomorphically to the higher areas.

To understand the nervous system as physical realization of the mind it is important, therefore,
  to think not in terms of neurons but to think in terms of structured nets.
This is the answer to our question Q1.
The permanent structure of the brain (Q2) is specified by neural connections
  and has the form of an overlay of net patterns each of which would,
  given time, stabilize itself under network self-organization.
Within cortical areas, nets are projected out of sets of neurons activated 
  by afferent excitation as coherent fields of neurons composed of net fragments.
Between cortical areas nets are coupled by homeomorphic fiber mappings.
In this way, brain-spanning nets are constructed. This is the answer to question Q3.
The connectivity structure that supports the generation of nets is generated (Q4)
  by network self-organization, the process that stabilizes connectivity patterns
  under self-interaction.

Brain-spanning nets that integrate sub-nets in all modalities 
  ---sensory, motor, langueage, emotions, representation of self etc. ---
  endow mental states with the coherence that has been posited
  as the essence of consciousness \cite{Kyoto,baars2005,tononi2012}.

\section{Conclusion}
The focus of this communication is the question how mental patterns are physically represented in the brain.
Fodor and Pylyshyn's discussion \cite{FodorPylyshyn} sets against each other the answers given by 
  the Classical and the Connectionist points of view.
This discussion laid bare a basic deficiency of neural models at the time,
  which they traced back to what later became known as the binding problem 
  \cite{Report,roskies1999binding,burwick2014binding}.
Their discussion simultaneously exposed the classical approach's
  total neglect of the origin of system structure.
Current machine learning systems are evidently overcoming these basic difficulties
  of the classical and the traditional neural approaches.
For the quest to understand mind and brain they, however,
   leave open important questions.
One of these is whether those systems could generate their mental patterns
  by direct interaction with the environment
  instead of inheriting them from human-generated training material,
  and another problem is how to interpret the vast random-looking arrays of numbers 
  that develop in those systems as mental patterns.
Even with the sole intention of better understanding machine learning systems
  and their possible fundamental limits it may be of value
  to formulate a cognitive architecture within the framework of what we know about brain and mind.
That is, what I am attempting here.
As central aspect, that architecture identifies mental patterns with what I call ``nets",
  neural connectivity patterns that are stable under self-interaction
  and in which each neuron's activity is supported and predicted by the activity of other neurons.
Nets as mental patterns solve the binding problem, have compositional structure as requested by F\&P,
  can be formed by self-organization under the influence of structured input
  and thus arise by emergence.

\bibliographystyle{splncs04}
\bibliography{Berlin23}

\end{document}